\documentclass[12pt,fleqn]{article}

\usepackage[T1]{fontenc}
\usepackage[ansinew]{inputenc}
\usepackage{amsmath}
\usepackage{indentfirst}
\usepackage{natbib}
\usepackage{graphics}
\usepackage{enumerate}

\begin{document}

\title{\bf FLUX DEPENDENCE OF PERSISTENT CURRENT\\ IN A MESOSCOPIC DISORDERED\\ TIGHT BINDING RING}
\author{J.Heinrichs\\Institut de Physique B5, University of Liege,\\
Sart-Tilman, B-4000 Liege, Belgium}
\date{}
\maketitle

\begin{abstract}
\thispagestyle{empty}

\noindent We reconsider the study of persistent currents in a disordered one-dimensional ring threaded by a magnetic flux $\phi$, using he one-band
tight-binding model for a ring of $N$-sites with random site energies.  The secular equation for the eigenenergies expressed in terms of transfer matrices in the
site representation is solved exactly to second order in a perturbation theory for weak disorder and fluxes differing from half-integer multiples of the flux
quantum $\phi_0=hc/e$.  From the equilibrium currents associated with the one-electron eigenstates we derive closed analytic expressions for the disorder
averaged persistent current for even and odd numbers,
$Ne$, of  electrons in the ground state.  Explicit discussion for the half-filled band case confirms that the persistent current is periodic with a period $\phi_0$,
as in the absence of disorder, and that its amplitude is generally suppressed by the effect of the disorder.  In comparison to previous results, based on an
approximate analysis of the secular equation, the current suppression by disorder is strongly enhanced by a new periodic factor proportional to
$1/\sin^2\frac{2\pi\phi}{\phi_0}$, for $\phi\neq\;(\text{integer})\;\phi_0/2$.\newline

\end{abstract}

\newpage
\pagestyle{plain}

\setcounter{page}{1}
\section{INTRODUCTION}
An Aharonov-Bohm flux threading a metallic or semiconducting ring leads to a persistent equilibrium current even if the ring is disordered, provided it is of
mesoscopic size smaller than the mean free path for inelastic scattering.  This is a direct consequence of quantum coherence which is not destroyed by elastic
impurity scattering.  The persisitent current is periodic  in the magnetic flux $\phi$ with a period given by the elementary flux quantum $\phi_0=hc/e$ (with $h$
the Planck~constant, $c$ the speed of light and -$e$ the electron charge).  The existence of persistent currents was predicted for one-dimensional rings by
BŸttiker, Imry and Landauer[1] and has later been observed experimentally in real three-dimensional rings by several groups[2-4].  Detailed theoretical
investigations of the persistent current have appeared for one-dimensional systems as well as for multichannel two- and three-dimensional rings\newline
[5-28]. 
Several reviews of the subject have also been published[29-31].\par

Many of these studies were motivated by the experiment of Levy {\it et al.}[2] whose result for the persistent current averaged over an ensemble of $10^7$
disordered copper rings is several orders of magnitude larger than the averaged current calculated for models taking only the scattering by random impurities into
account[14,16].  This led to attempts of in\-clu\-ding, in addition, the effect of the Coulomb interaction between the diffusing electrons[15,13,20,23,26], which
improved the agreement between measured and cal\-cu\-la\-ted persistent currents considerably.  However, in other recent numerical stu\-dies it was found that
Coulomb interactions suppress the persistent current in one-dimensional disordered systems, rather than enhancing it[21,24,25].\par

The present paper concerns the study of persistent currents in a one-dimensional disordered tight binding ring threaded by a magnetic flux, in the absence of
electron-electron interaction.  We recall that in previous discussions, starting with the key paper of BŸttiker {\it et al.}[1], the effect of the flux is described
via the flux-modified boundary condition (with $x$ the position along the one-dimensional ring of circumference $L$)

\begin{equation}\label{eq1}
\psi(x+L)=e^{i2\pi\frac{\phi}{\phi_0}}\psi(x)
\end{equation} 

\noindent for the ground state wavefunction of the system in the absence of flux[32,33].  This description of the flux has been demonstrated for the Schršdinger
equation in continuous space where the vector potential may be eliminated by choosing a gauge in which the external magnetic field vanishes in the interior of the
ring [32,33].\par 
We consider a tight-binding system composed of $N$ one-orbital atomic sites of spacing $a$ on a ring threaded by a magnetic flux in the direction perpendicular to
the ring.  This system is described by the set of difference equations

\begin{equation}\label{eq2}
-e^{i\frac{2\pi}{N}\frac{\phi}{\phi_0}}\varphi_{n+1}
-e^{-i\frac{2\pi}{N}\frac{\phi}{\phi_0}}\varphi_{n-1}
+\varepsilon_n\varphi_n=E\varphi_n\;,\;n=2,3,\ldots,N-1\quad ,
\end{equation} 

\begin{equation}\label{eq3}
-e^{i\frac{2\pi}{N}\frac{\phi}{\phi_0}}\varphi_2
-e^{-i\frac{2\pi}{N}\frac{\phi}{\phi_0}}\varphi_N
+\varepsilon_1\varphi_1=E\varphi_1\quad ,
\end{equation} 

\begin{equation}\label{eq4}
-e^{i\frac{2\pi}{N}\frac{\phi}{\phi_0}}\varphi_1
-e^{-i\frac{2\pi}{N}\frac{\phi}{\phi_0}}\varphi_{N-1}
+\varepsilon_N\varphi_N=E\varphi_N
\quad ,
\end{equation} 

\noindent which correspond to the familiar Anderson model for a disordered ring modified by the flux of the external vector potential[6].  Here $\varphi_n$ is the
amplitude of an eigenstate wavefunction at site $n$, $E$ and $\varepsilon_n$ are the corresponding eigenvalue and random site energies (potentials) in units of
minus a constant nearest-neighbour hopping parameter.\par

For clarity's sake we now briefly rederive the boundary conditions (1) for the discrete ring described by equations (2-4). To this end we perform a gauge
transformation

\begin{equation}\label{eq5}
\psi_n=e^{i\frac{2\pi}{N}\frac{\phi}{\phi_0}n}\varphi_n\;,\;n=1,2,\ldots, N\quad ,
\end{equation} 

\noindent to new amplitudes $\psi_n$, which transforms the system (\ref{eq2}-\ref{eq4}) into

\begin{equation}\label{eq6}
-\psi_{n+1}-\psi_{n-1}=(E-\varepsilon_n)\psi_n\;,\;n=2,3,\ldots, N-1\quad ,
\end{equation} 

\begin{equation}\label{eq7}
-\psi_2-e^{-i2\pi\frac{\phi}{\phi_0}}\psi_N=(E-\varepsilon_1)\psi_1\quad ,
\end{equation} 

\begin{equation}\label{eq8}
-e^{i2\pi\frac{\phi}{\phi_0}}\psi_1-\psi_{N-1}=(E-\varepsilon_N)\psi_N\quad ,
\end{equation} 

\noindent where the effect of the flux is transferred entirely to the bond linking the sites 1 and $N$.  These sites are bounding the continuous sequence of sites
$n=2,3,\ldots,N-1$ described by the equations \eqref{eq6} from which the flux is eliminated.  We now observe that the system (6-8) is formally equivalent to 

\begin{equation}\label{eq9}
-\psi_{n+1}-\psi_{n-1}=(E-\varepsilon_n)\;\psi_n,\; n=1,2,\ldots, N\quad ,
\end{equation} 

\noindent with the conditions (where (10.b) is the analog of (1) taken at the origin)

\begin{subequations}\label{eq10}
\renewcommand{\theequation}
{\theparentequation.\alph{equation}}\label{eq30}
\begin{align}
\psi_{N+1} &=e^{i\;2\pi\frac{\phi}{\phi_0}}\psi_1\quad ,\\
\psi_N &=e^{i\;2\pi\frac{\phi}{\phi_0}}\psi_0\quad ,
\end{align}
\end{subequations}

\noindent which describe a flux-free system defined by the $N$ equations (9), involving $N+2$ amplitudes $\psi_n$, in which the flux enters via the boundary
conditions (10.a,b).  Indeed, these boundary conditions ensure that the equations (9) for $n=1,N$ coincide with (7-8).  In addition, one sees that rotating all
amplitudes in (9) once around the ring ($\psi_m\rightarrow\psi_{m+N}, m=1,2,\ldots, N$) leaves these equations unchanged under the boundary condition

\begin{equation*}
\psi_{m+N}=e^{i\; 2\pi\frac{\phi}{\phi_0}}\psi_m\quad .\qquad\qquad\qquad\qquad\qquad\qquad\qquad\qquad\qquad\qquad\qquad\text{(10.c)}
\end{equation*}

\noindent This shows that the flux modified (or twisted) boundary condition (1) is also valid for the discrete ring.  We further recall that the similarity between
the boundary condition \eqref{eq1} and the pro\-per\-ty of Bloch functions in a periodic potential[1] (with $\frac{2\pi}{L}\frac{\phi}{\phi_0}$ playing the role of
the momentum and the ring size $L$ corresponding to the size of the unit cell) implies a flux periodicity with period $\phi_0$ of the eigenstates and
eigenenergies, and hence of the equilibrium persistent current carried by an energy level $E_n$,

\begin{equation}\label{eq11}
I_n=-c\frac{\partial E_n}{\partial\phi}\quad .
\end{equation} 

Cheung {\it et al.} [5] have studied the persistent current in the system described by equation (9) and (10.c) in terms of a two steps process: in the
first step the flux-free ring is being cut open and deformed int o a linear chain whose Schršdinger equation (9) is transformed to a tight-binding Bloch wave basis
[5,35] and is solved in terms of a transfer matrix for amplitudes of Bloch waves travelling to the right and to the left of the chain; in the second step the wave
amplitudes at the two ends of the chain are further matched by imposing the flux modified boundary condition (10.c).  This leads to an exact secular equation
for the eigenenergies of the ring involving only the flux and the total amplitude transmission coefficient, $t$, of the disordered chain of the first step at
energy $E$.  Cheung {\it et al.} combine their secular equation with (11) to calculate the persistent current, by making simple ad hoc assumptions about the
energy dependence of $t$ and approximating the effect of the disorder in the averaged transmission coefficient in terms of the Landauer~formula [5].  In the case of
weak disorder they find a constant, flux-independent, reduction of the amplitude of the persistent current with respect to its value for a perfect ring.\par

From a first-principles point of view the constant reduction of the current amplitude due to the disorder obtained by Cheung {\it et al.} [5] is surprising. 
 After all, the exact explicit expressions of the transmission- and reflection coefficients, $t$ and $r$, are complicated functions of the random site energies
$\varepsilon_j$ and of energy (through the dependence on the Bloch wavenumber, $k=\arccos(-E/2)$) [5,35], from which one may suspect the existence of a non
trivial interplay of disorder and flux in persistent currents in the exact eigenstates.\par

The purpose of this paper is precisely to discuss an exact first principles analytic calculation of the eigenenergies and persistent currents for the ring 
described by (2-4) in the framework of a second order perturbation theory for weak disorder.  In particular, we wish to study quantitatively the effect of the
expected interplay of disorder and magnetic flux on the magnitude of the persistent current.  On the other hand, a way of assessing the relative importance of
electron interaction effects in persistent currents [15,13,20,21,23-26] in the disordered regime, is by comparing numerical calculations with results obtained
for non-interacting electrons.  Exact analytical results for the effect of disorder on the persistent current for non interacting electrons, such as those
discussed in this paper, are clearly valuable in this context.\par

The perturbation theory for the eigenenergies of the ring for weak disorder is conveniently developed by starting from a secular equation expressed in terms of
transfer matrices for the Schršdinger
 equation (2-4) [or (6-8)] in the site representation.  The secular equation for the eigenenergies of the ring as well as their explicit determination to second
order in the random site energies are discussed in Sect.~II.  In Sect.~III we use the results of Sect.~II for calculating the persistent current averaged over the
disorder for a system of $N_e$ spinless particles in the ground state of the tight-binding system.  Our results will thus be relevant to average currents obtained
from measurements for a large number of weakly disordered mesoscopic rings corresponding to different realizations of the disorder[2].  The discussion of our
results for the typical low filling ($N_e\ll N$) and half-filled band ($N_e=\frac{N}{2}$) cases, their comparison with previous work and some concluding remarks
are presented in Sect.~IV.

\newpage
\section{TRANSFER MATRIX PERTURBATION \\THEORY}

Our aim is to calculate the electronic eigenenergies of the tight-binding ring described by (\ref{eq2}-\ref{eq4}) to successive orders in small random
fluctuations $\varepsilon_i$ of the site energies about a zero mean value.  This amounts to choosing the energy level of the individual atomic
sites as the zero of energy.  We find that the method based on tranfer matrices in the site-space representation[34,35] is ideally suited for this purpose since
these matrices depend linearly on the energies $\varepsilon_i$.\par

Defining the unimodular transfer matrices

\begin{equation}\label{eq12}
\hat P_n=
\begin{pmatrix}
-(E-\varepsilon_n)&-e^{-i\alpha}\\
e^{i\alpha}&0
\end{pmatrix},\;
\alpha=\frac{2\pi}{N}\frac{\phi}{\phi_0}\;,\;n=1,2,\ldots, N\quad ,
\end{equation} 

\noindent we have, from (\ref{eq2}-\ref{eq4}),

\begin{equation}\label{eq13}
\begin{pmatrix}
\varphi_{n+1}\\\varphi_n
\end{pmatrix}
=e^{-i\alpha}\hat P_n
\begin{pmatrix}
\varphi_n\\\varphi_{n-1}
\end{pmatrix}\;,\;
n=2,3,\ldots, N-1
\quad ,
\end{equation} 

\begin{equation}\label{eq14}
\begin{pmatrix}
\varphi_2\\\varphi_1
\end{pmatrix}
=e^{-i\alpha}\hat P_1
\begin{pmatrix}
\varphi_1\\\varphi_N
\end{pmatrix}\;,
\end{equation} 

\begin{equation}\label{eq15}
\begin{pmatrix}
\varphi_1\\\varphi_N
\end{pmatrix}
=e^{-i\alpha}\hat P_N
\begin{pmatrix}
\varphi_N\\\varphi_{N-1}
\end{pmatrix}\quad.
\end{equation} 

\noindent By combining  the amplitude vector of components $\varphi_N$ and $\varphi_{N-1}$, obtained by i\-te\-ra\-ting \eqref{eq13} in terms of the vector with
components $\varphi_2$ and $\varphi_1$, with (\ref{eq14}-\ref{eq15}), we readily obtain the following exact equation for the eigenvalues of the ring:

\begin{equation}\label{eq16}
\text{detm}\left(e^{-i2\pi\frac{\phi}{\phi_0}}\prod^N_{n=1}\hat P_n-\hat 1\right)=0
\quad,
\end{equation} 

\noindent which simplifies to

\begin{equation}\label{eq17}
\text{tr}\left(\prod^N_{n=1}\quad\hat P_n\right)=2\cos 2\pi\frac{\phi}{\phi_0}
\quad,
\end{equation} 

\noindent since the determinant of the product of unimodular matrices, $\prod^N_{n=1}\;\hat P_n$, is equal to unity.\par

For the purpose of our perturbation treatment for the eigenvalues for weak di\-sor\-der we first discuss the solutions of \eqref{eq17} in the absence of
disorder.\linebreak
In this case

\begin{equation}\label{eq18}
\prod^N_{n=1}\hat P_n\equiv\hat P^N,\;\text{with}\;\hat P=
\begin{pmatrix}
-E^0&-e^{-i\alpha}\\
e^{i\alpha}&0
\end{pmatrix}
\quad,
\end{equation}

\noindent where $E^0$ denotes eigenenergies of the perfect ring.  In order to determine the elements of the matrix $\hat P^N$ we diagonalize $\hat P$ by means of a
similarity transformation.  The eigenvalues of $\hat P$ are
\begin{equation}\label{eq19}
\lambda_{1,2}=\frac{1}{2}\left(-E^0\pm\sqrt{E^{0\;2}-4}\right)
\quad,
\end{equation}

\noindent  which may be rewritten in the form

\begin{equation}\label{eq20}
\lambda_{1,2}=e^{\pm i\;s}
\quad ,
\end{equation}

\noindent with

\begin{equation}\label{eq21}
E^0=-2\cos s
\quad .
\end{equation}

\noindent The similarity transformation which diagonalizes $\hat P$ is defined by the non-unitary eigenvector matrix

\begin{equation}\label{eq22}
\hat U=
\begin{pmatrix}
e^{i(s-\alpha)}&e^{-i(s+\alpha)}\\
1&1
\end{pmatrix},\;
\text{with}\; \hat U^{-1}
=\frac{e^{i\alpha}}{2i\sin s}
\begin{pmatrix}
1&-e^{-i(s+\alpha)}\\
-1&e^{i(s-\alpha)}
\end{pmatrix}\quad .
\end{equation}

\noindent Using the fact that $\left(\hat U^{-1}\hat P\hat U\right)_{ij}=\lambda_i\delta_{i,j},\;i=1,2$, we then obtain from \eqref{eq21} and \eqref{eq22},

\begin{equation}\label{eq23}
\hat P^m=\hat U\left(\hat U^{-1}\hat P\hat U\right)^m\hat U^{-1}
=\frac{1}{\sin s}
\begin{pmatrix}
\sin(m+1)s&-e^{-i\alpha}\sin ms\\
e^{i\alpha}\sin ms&-\sin(m-1)s
\end{pmatrix}
\quad .
\end{equation}

\noindent In the absence of disorder the eigenvalue equation \eqref{eq17} then reduces to

\begin{equation}\label{eq24}
\cos Ns=\cos\frac{2\pi\phi}{\phi_0}
\quad ,
\end{equation}

\noindent which leads to the energy levels

\begin{equation}\label{eq25}
E^0_k(\phi)=-2\cos\frac{2\pi}{N}\!\left(k+\frac{\phi}{\phi_0}\right)
\quad ,
\end{equation}

\noindent using \eqref{eq21}, and to the corresponding equilibrium currents (equation \eqref{eq11})

\begin{equation}\label{eq26}
I^0_k=-\frac{2e}{N\hbar}\sin\frac{2\pi}{N}\!\left(k+\frac{\phi}{\phi_0}\right)
\quad ,
\end{equation}

\noindent where $k=0,\;\pm 1,\;\pm2,\ldots$.  These results coincide with those obtained earlier by Cheung {\it et al.}[5] for the perfect tight-binding ring.  For
levels with $|k|\ll N$ the energies \eqref{eq25} as a function of $\phi/\phi_0$ grow quadratically with $\phi/\phi_0$ away
from minima at $\phi/\phi_0=-k,\;k=0,\;\pm1,\ldots$.  For such levels therefore, the energy spectrum is similar to the familiar free particle spectrum composed of
intersecting parabolas whose ordinates increase as a function of
$\phi/\phi_0$ away from their respective vertices at $0,\;\pm1,\;\pm 2,\ldots$[5].\par

We proceed with the solution of \eqref{eq17} for weak random fluctuations of the site energies where we expand the equation to successive orders in the
$\varepsilon_i$ and obtain the corresponding first and second orders corrections, $E^{(1)}_k$ and $E^{(2)}_k$, for the perturbed energy levels,

\begin{equation}\label{eq27}
E_k=E^0_k+E^{(1)}_k+E^{(2)}_k+\cdots,\;k=0,\;\pm 1,\ldots
\quad .
\end{equation}

\noindent The transfer matrices $\hat P_n$ are split into unperturbed and perturbed parts:

\begin{equation}\label{eq28}
\hat P_n=\hat P+\hat V_n\equiv\hat P-(\Delta E-\varepsilon_n)
\begin{pmatrix}
1&0\\0&0
\end{pmatrix}
\quad ,
\end{equation}

\noindent where

\begin{equation}\label{eq29}
\Delta E=E^{(1)}+E^{(2)}+\cdots
\quad ,
\end{equation}

\noindent and the matrix product, $\prod^N_{n=1}\hat P_n$, is expanded to quadratic order in the quantities $\Delta E-\varepsilon_n$.  This yields

\begin{multline}\label{eq30}
\prod^N_{n=1}\hat P_n=\hat P^N+\sum^N_{m=1}\hat P^{m-1}\;\hat V_m\;\hat P^{N-m}\\
+\sum^N_{n=2}\sum^{n-1}_{m=1}\hat P^{m-1}\;\hat V_m\;\hat P^{n-m-1}\;\hat V_n\;\hat P^{N-n}+\cdots
\quad .
\end{multline}

\noindent By inserting this expression in \eqref{eq17}, using \eqref{eq23}, and equating the two sides of the equation order by order we obtain, after some
straightforward algebra,

\begin{equation}\label{eq31}
E^{(1)}_k=E^{(1)}=\frac{1}{N}\sum^N_{n=1}\varepsilon_n
\quad ,
\end{equation}
\noindent and

\begin{multline}\label{eq32}
E^{(2)}_k=\frac{1}{N\sin q_k\sin Nq_k}\sum^N_{n=2}\sum^{n-1}_{m=1}
(E^{(1)}-\varepsilon_m)(E^{(1)}-\varepsilon_n).\\
\sin (n-m) q_k . \sin (N-n+m) q_k
\quad ,
\end{multline}

\noindent where

\begin{equation}\label{eq33}
q_k=\frac{2\pi}{N}\left(k+\frac{\phi}{\phi_0}\right)
\quad .
\end{equation}

\noindent Clearly, the weak disorder perturbation theory breaks down at flux values equal to half-interger multiples of the flux quantum $\phi_0$ where
\eqref{eq32} diverges.  Such divergencies are familiar[19] in other perturbative studies of persistent currents[13,14,16].

\newpage
\section{AVERAGED PERSISTENT CURRENT}

The random eigenenergy corrections \eqref{eq31} and \eqref{eq32} are easiest to analyse through low order moments.  Here we shall consider only the averaged
ei\-gen\-e\-ner\-gies and the corresponding averaged persistent currents.  These are relevant to experiments obtaining the persistent currents in a large number of
identical mesoscopic rings corresponding to different realizations of the disorder[2].  In averaging the eigenenergies \eqref{eq27} we assume a site energy
disorder in the form of gaussian white noise with a correlation

\begin{equation}\label{eq34}
\langle \varepsilon_i\;\varepsilon_j\rangle=\varepsilon_0^2\;\delta_{i,j}
\quad .
\end{equation}

\noindent From \eqref{eq31} and \eqref{eq32} we thus obtain $\langle E^{(1)}_k\rangle=0$ and

\begin{equation}\label{eq35}
\langle E^{(2)}_k\rangle=-\frac{\varepsilon^2_0}{N^2\sin q_k\;\sin Nq_k}\sum^N_{n=2}\sum^{n-1}_{m=1}
\sin (n-m)q_k\sin (N-n+m)q_k
\quad .
\end{equation}

\noindent By performing the finite summations over sites in \eqref{eq35} we finally get

\begin{equation}\label{eq36}
\langle E^{(2)}_k\rangle=\frac{\varepsilon^2_0}{4\sin q_k}\left[\left(1+\frac{1}{N}\right)\cot Nq_k-\frac{1}{N}\cot q_k\right],\;
k=0,\;\pm 1,\;\pm2,\ldots
\quad .
\end{equation}

\noindent From \eqref{eq11}, \eqref{eq25} and \eqref{eq36} we then obtain the averaged current from the level $E_k$

\begin{multline}\label{eq37}
\langle I_k\rangle=-\frac{e}{N\hbar}\left\{2\sin q_k-\frac{N\varepsilon^2_0}{4\sin q_k}\cdot\right.\\
\left.\left[\left(1+\frac{1}{N}\right)\left(\frac{1}{\sin^2\frac{2\pi\phi}{\phi_0}}+\frac{1}{N}\cot\frac{2\pi\phi}{\phi_0}\cot\;q_k\right)
+\frac{1}{N^2}\left(1+2\cot^2q_k\right)\right]\right\}
\quad ,
\end{multline}

\noindent which reduces to

\begin{multline}\label{eq37a}
\langle I_k\rangle=-\frac{e}{N\hbar}\left[2\sin\frac{2\pi}{N}\!\left(k+\frac{\phi}{\phi_0}\right)
-\frac{N\varepsilon^2_0}{4\sin^2\frac{2\pi\phi}{\phi_0}}\;\frac{1}{\sin\frac{2\pi}{N}\!\left(k+\frac{\phi}{\phi_0}\right)}\right],\\
\;k=0,\;\pm 1,\ldots
\quad ,
\end{multline}

\noindent by dropping terms of order $\frac{1}{N}$ and $\frac{1}{N^2}$ in the square brackett of \eqref{eq37}.\par

In order to obtain the total averaged persistent current $\langle I\rangle$ for a system of $N_e$ electrons in the ground state at a given $\phi/\phi_0$ we must
sum the contributions from the lowest occupied levels at the considered flux value.  For simplicity we ignore spin and assume that a level may be occupied only by
one electron.  As is well known the current is different for even and for odd numbers of electrons[5].  We first concentrate on the contribution to the total
current from the zeroth order levels corresponding to the non-disordered system.  For odd $N_e$ and for values of flux in the interval
$-\frac{1}{2}\leq\frac{\phi}{\phi_0}<\frac{1}{2}$ this contribution, $I^{(0)}$, is obtained by summing the one-electron currents \eqref{eq26} from
$k=-\frac{N_e-1}{2}\;\text{to}\; k=\frac{N_e-1}{2}$.  The fact that these values of $k$ correspond to the $N_e$ lowest levels of the spectrum in the considered
range of fluxes is easily verified e.g.  in the free particle-like limit of \eqref{eq25} where the eigenernergies reduce to intersecting parabolas whose vertices
on the
$\phi$-axis correspond to integer values of $\phi/\phi_0$.  One finds in this case[5]

\begin{equation}\label{eq38}
I^{(0)}=\sum^{(N_e-1)/2}_{k=-(N_e-1)/2}\;I^0_k=-I_0\frac{\sin\frac{2\pi\phi}{N\phi_0}}{\sin\frac{\pi}{N}}\simeq-2\frac{\phi}{\phi_0}I_0
\quad ,
\end{equation}

\noindent where

\begin{equation}\label{eq39}
I_0=\frac{e\;v^b_F}{N},\;\text{with}\;v^b_F=\frac{2}{\hbar}\sin\frac{\pi\;N_e}{N}
\quad .
\end{equation}

\noindent The meaning of the velocity $v^b_F$ is clear: its value coincides with the quantity $\frac{N}{h}\frac{\partial E^0_k}{\partial k}\bigg|_{k=N_e/2}$ which
is the band velocity 	at an energy just above the highest occupied level, that is a velocity close to the velocity at the fermi level.  It may also be simply
related to a fermi velocity as follows.  Consider a non-disordered linear tight-binding chain whose electronic states are described by the equation
$-V(\varphi_{n+1}+\varphi_{n-1})=E\;\varphi_n$ for a system of $N$ sites.  In the continuum limit \newline
($a\rightarrow 0$) this equation becomes (with $x=na$) 
$-a^2Vd^2\varphi(x)/dx^2=(E-2V)\varphi(x)$, which is the Schršdinger equation for a free particle of mass $m=\hbar^2/(2Va^2)$.  In this limit the fermi velocity
for a system of $N_e$ electrons is \linebreak
$v_F=\hbar k_F/n=\frac{2V}{\hbar}\frac{N_e\pi}{N}$ (with $k_F=N_e\pi/(Na)$) the fermi wavenumber).  This shows that
$v_F$ is the limiting value of the band velocity in \eqref{eq39} for low band filling, \linebreak $N_e\pi\ll N$.
This discussion shows that $I_0$ is the current associated with a single electron of velocity $v^b_F$ at the fermi level (or of fermi velocity $v_F$ for low band
filling).  This is the well-known manifestation of the strong compensation of the contributions to the persistent current from successive occupied levels of the
system, with the result that the total current is approximately equal to the current from the highest occupied level.\linebreak

Similarly, from the structure of the level spectrum \eqref{eq25}, it follows that for even $N_e$ states for $1\leq\phi/\phi_0<0$ the $N_e$ lowest
occupied states correspond to \linebreak
$k=0,\;\pm 1,\;\pm 2,\ldots,\;\pm \left(\frac{N_e}{2}-1\right),\;-\frac{N_e}{2}$.  In this case one thus obtains[5]

\begin{equation}\label{eq40}
I^{(0)}=-I_0\frac{\sin\left[\frac{\pi}{N}\!\left(\frac{2\phi}{\phi_0}-1\right)\right]}{\sin\frac{\pi}{N}}
\simeq-\left(\frac{2\phi}{\phi_0}-1\right)\;I_0
\quad .
\end{equation}

Obviously the strong compensation of contributions from adjacent levels to the pure system current shown by (\ref{eq39}-\ref{eq40}) does not occur when summing the
effect of the disorder in \eqref{eq37a} for the successive occupied levels.  Thus, unlike for the contribution from the non-disordered energy levels, we expect to
obtain a reasonable estimate of the effect  of the disorder on the averaged current by approximating the discrete sum over occupied levels by an integral.  More
precisely, we shall use the Euler-MacLaurin summation formula[36]:

\begin{equation}\label{eq41}
\sum^{n-1}_{k=1}f_k\simeq \int^n_0f(k)dk-\frac{1}{2}\left[f(0)+f(n)\right]\quad ,
\end{equation}

\noindent where we have included only the leading integrated terms[36].\par

For odd $N_e$ and $-\frac{1}{2}
\leq\frac{\phi}{\phi_0}<\frac{1}{2}$ where the summation over occupied levels runs over $k=0,\;\pm 1,\ldots,\;\pm
\left(\frac{N_e-1}{2}\right)$ we obtain, by applying \eqref{eq41},

\begin{equation}\label{eq42}
\sum_k\frac{1}{\sin\frac{2\pi}{N}\!\left(k+\frac{\phi}{\phi_0}\right)}
=-\frac{1}{2}\left(\frac{1}{\sin r_+}-\frac{1}{\sin r_-}\right)
-\frac{N}{4\pi}\text{ln}\;\frac{(1+\cos r_+)(1-\cos r_-)}{(1-\cos r_+)(1+\cos r_-)}
\quad ,
\end{equation}

\noindent where

\begin{equation}\label{eq43}
r_\pm=\frac{\pi}{N}\left(N_e+1\pm 2\frac{\phi}{\phi_0}\right)
\quad .
\end{equation}

\noindent In the case of even $N_e$, for $0\leq\phi/\phi_0<1$, the summation over $k$ is over the values $k=0,\;\pm 1,\ldots,\;\pm\left(\frac{N_e}{2}-1\right)$
and $-\frac{N_e}{2}$ and we find

\begin{equation}\label{eq44}
\sum_k\frac{1}{\sin\frac{2\pi}{N}\!\left(k+\frac{\phi}{\phi_0}\right)}
=-\frac{1}{2}\left(\frac{1}{\sin s_+}+\frac{1}{\sin s_-}\right)
-\frac{N}{4\pi}\text{ln}\;\frac{(1+\cos s_+)(1-\cos s_-)}{(1-\cos s_+)(1+\cos s_-)}
\quad ,
\end{equation}

\noindent where

\begin{equation}\label{eq45}
s_\pm=\frac{\pi}{N}\left(N_e\pm 2\frac{\phi}{\phi_0}\right)
\quad .
\end{equation}

\noindent After some simplification of the disorder contributions the expressions for the total averaged current, $\langle I\rangle=\sum_k\langle I_k\rangle$ given
by \eqref{eq37a}, \eqref{eq38} and (\ref{eq42}-\ref{eq43}) for odd $N_e$ and by \eqref{eq37a}, \eqref{eq40} and (\ref{eq44}-\ref{eq45}) for even $N_e$,
respectively, reduce finally to (with $k_F=\frac{\pi\,N_e}{Na}$)

\begin{multline}\label{eq46}
\langle I\rangle=
-I_0
\left\{\frac{\sin\frac{2\pi\phi}{N\phi_0}}{\sin\frac{\pi}{N}}
      +\frac{N\;\varepsilon^2_0}{8\sin^2\frac{2\pi\phi}{\phi_0}\sin k_Fa}.\right.\\  
\left.\left[\frac{\cos(k_Fa+\frac{\pi}{N})   \sin\frac{2\pi\phi}{N\phi_0}}{\sin^2\frac{2\pi\phi}{N\phi_0}-\sin^2\left(k_Fa+\frac{\pi}{N}\right)}
+\frac{N}{2\pi}
\text{ln}\;\left(\frac{\sin\left(k_Fa+\frac{\pi}{N}\right)-\sin\frac{2\pi\phi}{N\phi_0}}{\sin\left(k_Fa+\frac{\pi}{N}\right)+\sin\frac{2\pi\phi}{N\phi_0}}\right)\right]\right\}\\
\text{for odd}\;N_e, -\frac{1}{2}<\frac{\phi}{\phi_0}<\frac{1}{2},\;\phi\neq 0
\quad ,
\end{multline}

\noindent and

\begin{multline}\label{eq47}
\langle I\rangle=
-I_0
\left\{\frac{\sin\left[\frac{\pi}{N}\!\left(\frac{2\phi}{\phi_0}-1\right)\right]}{\sin\frac{\pi}{N}}
      -\frac{N\;\varepsilon^2_0}{8\sin^2\frac{2\pi\phi}{\phi_0}\sin \frac{\pi\;N_e}{N}}.\right.\\  
\left.\left[\frac{\sin k_Fa\cos\frac{2\pi\phi}{N\phi_0}}{\sin^2\frac{2\pi\phi}{N\phi_0}-\sin^2 k_Fa}
-\frac{N}{2\pi}
\text{ln}\;\left(\frac{\sin k_Fa-\sin\frac{2\pi\phi}{N\phi_0}}{\sin k_Fa+\sin\frac{2\pi\phi}{N\phi_0}}\right)\right]\right\}\\
\text{for even}\;N_e, 0<\frac{\phi}{\phi_0}<1, \phi\neq\frac{1}{2}
\quad .
\end{multline}

\noindent The formulae (\ref{eq46}-\ref{eq47}) together with \eqref{eq32} and \eqref{eq37} are the main results of the present paper.

\newpage
\section{DISCUSSION AND CONCLUDING REMARKS}

In discussing the averaged persistent currents \eqref{eq46} and \eqref{eq47} we consider successively two~limits for the so-called band filling factor $N_e/N$: the
case of low filling,$\frac{\pi\;N_e}{N}\ll1,\; N_e\gg1$ , and the half-filled band case, $N_e=N/2$, for even $N$.  For $\pi\;N_e\ll N$, one easily verifies that the
disorder corrections in (\ref{eq46}-\ref{eq47}) are of opposite signs to those of the perfect systems currents i.e. the disorder suppresses the persistent
current.  On the other hand, in the half-filled band case, $N_e=\frac{N}{2}$, we obtain successively from \eqref{eq46} and \eqref{eq47}

\begin{multline}\label{eq48}
\langle I\rangle=
-I_0\frac{2\phi}{\phi_0}\left(1-\frac{N\;\varepsilon^2_0}{8\sin^2\frac{2\pi\phi}{\phi_0}}\right),\\
\text{for odd}\; N_e, \; -\frac{1}{2}<\frac{\phi}{\phi_0}<\frac{1}{2},\;\phi\neq 0
\quad ,
\end{multline}

\noindent and

\begin{multline}\label{eq49}
\langle I\rangle=
-I_0\left(\frac{2\phi}{\phi_0}-1\right)\left(1-\frac{N\;\varepsilon^2_0}{8\sin^2\frac{2\pi\phi}{\phi_0}}\right),\\
\text{for even}\; N_e, \; 0<\frac{\phi}{\phi_0}< 1,\;\frac{\phi}{\phi_0}\neq \frac{1}{2}
\quad ,
\end{multline}

\noindent to leading order in $\frac{2\pi}{N}\frac{\phi}{\phi_0}$.  The equation (\ref{eq48}-\ref{eq49}) show that for the half-filled band cases the effect of the
disorder is simply to multiply the perfect system current by a pe\-ri\-o\-dic flux dependent reduction factor of period $\phi_0/2$.  This reduction factor does not
alter the periodicity with period $\phi_0$ of the original perfect system current discussed in[5-7].  We recall that the reduction of the current due to
the disorder has a simple interpretation in terms of level repulsion caused by the random impurity potential.  Consider, for simplicity, the limit where the
conducting electrons are free and the energy level spectrum as a function of
$\phi/\phi_0$ is composed of intersecting parabolas.  As is well-known, the one-body impurity potential causes level repulsion which leads to the replacement of
the free parabola crossings by avoided crossings.  The latter are more flat than the original true crossings, which leads to a reduction of the currents $\langle
I_k\rangle=-c\partial\langle E_k\rangle/\partial\phi$ near them, caused by the interaction with the impurities.\par

The expression (\ref{eq48}-\ref{eq49}) for weak randomness
differ remarkably from the corresponding results of by Cheung {\it et al.}[5].  While our results are obtained from an exact solution of the secular equation (17) 
to second order in the random site energies for weak randomness, those of Cheung {\it et al.} [5] follow from an approximate study of their exact eigenvalue
equation

\begin{equation}\label{eq51}
Re\left(\frac{1}{t}\right)=\cos 2\pi\frac{\phi}{\phi_0}\quad ,
\end{equation}

\noindent obtained from (9) and (10.c) after transforming to a Bloch wave representation as discussed in Sect. I.  We recall that $t$ is the complex transmission
 coefficient at energy $E$ across a disordered linear chain of length $Na$ obtained by cutting the ring open.  From simple but crude assumptions about the
energy dependence of $t$ Cheung {\it et al.} derive an expression for the current from a level $E_k$ which may be approximated by $I_k\simeq I_0| t|$
[5].  For a half-filled band, assuming that the total current in the presence of disorder is approximately equal to $-(2\phi/\phi_0\;-1)$ times the current carried
by the last occupied level they obtain[37]

\begin{equation}\label{eq52}
\langle I\rangle=
-\left(\frac{2\phi}{\phi_0}-1\right)\;I_0\langle|t|\rangle
\quad ,
\end{equation}

\noindent where $\langle|t|\rangle$ is the averaged transmission coefficient measured at the fermi level which may be related to the localization leength by using
the Landauer formula.  For weak disorder $\langle|t|\rangle$ is thus found to decay as
$N\;\varepsilon^2_0$ (in our notations[38]), with a constant coefficient,  from the value one for the ordered system (See Sect.~III.C of [5]).  In contrast the
exact weak disorder effect in
\eqref{eq49}, which is also proportional to $N\varepsilon^2_0$, varies strongly with the flux, as $1/\sin^2\frac{2\pi\phi}{\phi_0}$.  The strong flux dependence of
the disorder effect in the persistent current (\ref{eq48}-\ref{eq49}) shows that the approximations used by Cheung {\it et al.} [5] in solving \eqref{eq51} are
insufficient, at least for the case of weak disorder.\par

In discussing the magnitude of the effect of the disorder in (\ref{eq48}-\ref{eq49}) relative to the perfect system current, it is useful to recall the domain of
validity of these perturbative expressions.  This domain is determined by:

\begin{enumerate}
\item the condition, $|\varepsilon_0|\ll 1$, of smallness of the amplitude fluctuations of the site energies relative to the hopping parameter in
(\ref{eq2}-\ref{eq4}) 
\item the inequality $\sqrt{\langle E^{(1)\;2}\rangle}\ll\Delta$, which expresses that the typical shift of energy levels due to disorder must be small compared to
the spacing $\Delta$ of energy levels at zero flux at the fermi surface[5,7].  From \eqref{eq25} we have $\Delta=\frac{4\pi}{N}.\sin k_Fa$, and by using
\eqref{eq31} and \eqref{eq34} this condition reduces to

\begin{equation}\label{eq52}
\sqrt N|\varepsilon_0|\ll 4\pi
\quad ,
\end{equation}
\end{enumerate}

\noindent  in the half-filled band case.  This implies that within the perturbative domain the disorder term in the brackett of \eqref{eq49} may be enhanced to
values of order unity by the factor $1/\sin^2\frac{2\pi\phi}{\phi_0}$, for fluxes sufficiently close to multiples of $\phi_0/2$.\linebreak
For odd $N_e$ this leads to a strongly enhanced suppression of the current near $\phi=\pm\frac{\phi_0}{2}$ while for even $N_e$ a similar enhanced suppression
occurs near $\phi=0$ and near $\phi=\pm\phi_0$.  Such large effects of a weak disorder may exist only for mesoscopic rings since e.g. for $|\varepsilon_0|=0.1$
values of $N$ compatible with \eqref{eq52} would be restricted to $N\sim10^2$.\par

We conclude by suggesting that important modifications of the amplitude of persistent currents in discrete rings due to the interplay of disorder and magnetic flux
may also exist for more realistic multichannel disordered rings as well as for models incorporating Coulomb interaction.  In view of the persisting disagreements
between calculated and measured persistent currents in metallic rings (see Sect.~ÊI) such effects would seem to deserve further detailed study.


\begin{thebibliography}{99}
\bibitem{38}The relation between the rectangular distribution of width $W$ for the site energies used in [5] and the gaussian parameter $\varepsilon_0$ is
obtained by identifying the second moment of this distribution with $\varepsilon^2_0=\langle\varepsilon^2_j\rangle,\; j=1,\ldots,\; N$.  This yields
$\varepsilon^2_0=W^2/12$.
\end{thebibliography}
\end{document}